\documentclass[11pt,reqno]{amsart}
\usepackage{graphicx}
\usepackage{amssymb,amscd,amsbsy}
\usepackage{amsthm}

\usepackage{amsmath}
   \usepackage{amsfonts}   
   \usepackage{amssymb}

\usepackage{graphicx}
\usepackage{amsmath}

\newcommand{\noi}{\noindent}

\def\Rat{\text{Rat}\,}

\def\d{\delta}

\def\MM{\mathcal   M}
\def\HH{\mathcal   H}

\def\G{\Gamma}

\def\l{\lambda}

\def\[{\left[}
\def\]{\right]}
\def\({\left(}
\def\){\right)}

\newcommand{\C}{{\mathbb C}}

\newcommand{\p}{\partial}
\newcommand{\eeq}{\end{equation}}
\newcommand{\beq}{\begin{equation}}
\newcommand{\bay}{\begin{eqnarray}}
\newcommand{\ey}{\end{eqnarray}}
\newcommand{\bey}{\begin{eqnarray*}}
\newcommand{\eey}{\end{eqnarray*}}

\newcommand{\R}{\operatorname{res}}

\newcommand{\tr}{\operatorname{trace}}

\newtheorem{thm}{\hspace{\parindent}Theorem}[section]

\newcommand{\CP}{{\mathbb C}{\mathbb  P}}

\newcommand{\pa}{\partial}
\theoremstyle{remark}
\newtheorem*{rem*}{Remark}

\usepackage{amsfonts,amssymb}
\usepackage{amsthm}
\usepackage{amscd}
\usepackage{array}
\usepackage{color}
\usepackage{pstricks,pst-node}
\usepackage{pst-plot}
\usepackage{pst-solides3d}

\begin{document}

\newcommand{\vse}{\vspace{.2in}}
\numberwithin{equation}{section}

\title{Infinite hierarchies of Poisson structures for  integrable systems and spectral curves.}
\author{K.  Vaninsky}
\begin{abstract}
In this short survey  we describe our approach for constructing    hierarchies of Poisson brackets for classical  integrable systems using its'  spectral curves.
\end{abstract}
\maketitle

\tableofcontents
\setcounter{section}{0}
\setcounter{equation}{0}

\section{Introduction.} Hamiltonian formalism is the main mathematical apparatus of the classical and quantum physics. Hamiltonian formalism of completely integrable systems was intensively studied in the last  fifty years. Starting with pioneering work of   work of Gardner, \cite{G}, and Zakharov and Fadeev, \cite{ZF},  different  approaches were developed  by  Leningrad school of Faddeev, \cite{FT},  Belavin and Drinfeld \cite{BD},  Gelfand-Dikey, \cite{GD},  Attiyah--Hitchin   \cite{AH},  Magri, \cite{MA},  Novikov and Veselov \cite{NV}, Krichever and Phong, \cite{KP}, with hundreds papers following these paths.  Nevertheless the subject remains incomplete, there is no approach that covers all known examples integrable by the methods of spectral curves and algebraic geometry. Even the simplest questions have no answer.  The ongoing work of the author   in an attempt to fill this gap.

Let us describe a general framework.  It is believed that on  phase space $\MM$ of completely integrable system there exists  a  finite or  infinite set of
commutative vector fields
$$
X_1, X_2,\hdots
$$
that are compatibility conditions for the commutator representation or  Lax's equations
$$
X_k:\qquad\qquad L=\[L,A_k\],\qquad\qquad k=1,2,\hdots.
$$
This vector fields can be written  with  the classical Poisson bracket $\{\;, \; \}_{\pi_0}$
and different      Hamiltonians   $\HH_1, \HH_2, \HH_3,...;$  as
$$
X_k=\{\;\;,\HH_k\}_{\pi_0},\qquad k=1,2,....
$$
It is believed  that on the  phase space $\MM$  there exists an infinite  sequence of compatible Poisson brackets
\beq\label{ihpb}
\{\;,\;\}_{\pi_0}\qquad  \{\;,\;\}_{\pi_1}\qquad  \{\;,\;\}_{\pi_2}\qquad .....
\eeq
Any vector field  of  the   hierarchy can be written using  these brackets and different Hamiltonians  $\HH_0, \HH_1, \HH_2,...$
$$
X_k=\{\;\;,\HH_{k-p}\}_{\pi_p}, \qquad k=1,2,.... ;\qquad  0\leq p \leq k.
$$

For each integrable system an explicit form of the Poisson brackets \ref{ihpb} on $\MM$ is  different. The first few usually can be written down explicitly but the  formulas quickly  become unmanageable.

\section{The main conjecture.}
In our approach we  construct a parametrization
of the space $\MM$   in terms  of  the Hurwitz space
\beq\label{dst}
\MM\longrightarrow (\G,dp, dE,  \chi).
\eeq
The Riemann surface $\G$ with two differentials  $dp$ and $dE$ arises for the periodic spectral for the operator $L$ or its two-dimensional analogs
entering the commutator formalism, \cite{KR}.  Moreover, on $\G$ there exist a meromorphic function  $\chi(Q)=\chi(x,Q)$ which we call the Weyl function on the Riemann surface. It is closely connected with the standard Weyl function which arises in classical spectral theory, \cite{MaV},  and to the classical Baker--Akhiezer function $e(x,Q)$ by the formula (for the KdV equation)
$$
i\chi(x,Q)= \frac{\partial}{\partial x} \log e(x,Q).
$$
The careful description of the image of this {\it direct spectral transform} is required but this can be done for the most interesting examples. The quadruples
$(\G,dp, dE,  \chi)$ are the main object of our consideration.

Within our approach  we write  Poisson brackets  in  form
\beq\label{apb}
\{\chi(P),\chi(Q)\}^f= \sum \int\limits_{\overset{\curvearrowright}{O}_k} \omega_{P\, Q}^f\; ,
\eeq
where  the evaluation map is defined as
$$
Q:\; (\Gamma, \chi)\rightarrow \chi(Q),\qquad\qquad\qquad Q\in \Gamma.
$$
The  meromorphic one-form $\omega_{PQ}^f$ depends on the functional parameter $f$.   The one-form has poles at the poles of $\chi$,  at the points $P, Q$  and at infinities of $\G$. The small circles $O_k$ surround  poles of $\chi$. The sum can be finite or infinite depending on the total number of poles of $\chi$.

We conjecture that a single formula \ref{apb} with different choices of the meromorphic one--form $\omega_{PQ}$ describes infinite hierarchies  \ref{ihpb} of Poisson brackets. This is not a theorem but a guiding principle which we checked for a few examples  described bellow.

\section{The  Camassa--Holm  equation.}
The Camassa--Holm equation    is an approximation to the Euler equation
describing  an ideal fluid
$$
X_1:\qquad\qquad {\p v\over \p t} + v{\p v\over \p x} +{\p  \over \p x}R\[v^2 +{1\over 2} \({\p v\over \p x}\)^2\]=0$$
in which $t\geq 0$ and $  -\infty < x < \infty$, $v=v(x,t)$ is velocity, and $R$ is inverse to $L=1-d^2/d x^2$ {\it i.e.}
$$
R[ f] (x)= {1\over 2} \int\limits_{-\infty}^{+\infty} e^{-|x-y|} f(y) dy.
$$
Introducing the function $m=L[v]$ one  writes the equation in the form
\footnote{We use   notation $D$ for the $x$-derivative and  $\bullet$ for the $t$--derivative. We use $\delta$ for the Frechet derivative.}
$$
m^{\bullet}+\(mD+Dm\)v=0.
$$
The CH equation is a Hamiltonian system
$m^{\bullet} +\{m,\HH_1\}_{\pi_0}=0
$
with  Hamiltonian
$$
\HH_1={1\over 2}\int_{-\infty}^{+\infty} mv\, dx=\text{energy}
$$
and the bracket
\beq\label{ppb}
\{A,B\}_{\pi_0}=\{A,B\}=\int_{-\infty}^{+\infty}{\delta A\over \delta m} \(mD+Dm\) {\delta B\over \delta m}\, dx.
\eeq
We consider the CH equation with nonnegative ($m\geq 0$) initial data and such   decay at infinity that:
$$
\int_{-\infty}^{+\infty} m(x) e^{|x|} dx < \infty.
$$
We denote this class of functions by $\MM$.
One can associate to the CH equation an auxiliary string spectral problem,
\footnote{We use prime $'$ to denote $\xi$-derivative.}
$$
f''(\xi)+ \l g(\xi)f(\xi)=0,\qquad\qquad -2\leq \xi \leq 2.
$$
The background information for  this spectral problem can be found in  \cite{GK, KK2, DM}.
The variables $\xi$ and $x$ are related by
$$
x\longrightarrow \xi =2\tanh {x\over 2}.
$$
Also the potential $g(\xi)$ is related to $m(x)$ by the formula $g(\xi)=m(x)\cosh^4{x\over 2}$.
For initial data from $\MM$ the total mass of the associated string is finite $\int_{-2}^{+2} g(\xi)d\xi <\infty$.
For simplicity we consider $N$-peakon solutions of the CH equation. In this case $g(\xi)$ is a sum of finite number of point masses.

Two important solutions $\varphi(\xi,\l)$ and $\psi(\xi,\l)$ of the string spectral problem  are specified by   initial data
\bey
\varphi (-2,\l)& =1\qquad\qquad\qquad \;   \psi(-2,\l)=0\\
\varphi'(-2,\l)&=0\qquad\qquad\qquad      \psi'(-2,\l)=1.
\eey
The  Weyl function is defined by the formula
$$
\chi(\l)=-\frac{\varphi(2,\l)}{\psi(2,\l)}.
$$
The Riemann surface $\Gamma$ associated with the string spectral problem consists of two components
$\Gamma_+$ and $\Gamma_-$ two copies of the Riemann sphere. The points $\l$'s where two spheres are glued to each other are  points of the Dirichlet spectrum. The points $\gamma$'s on $\Gamma_-$ are the points of the Newmann spectrum, see Figure 1.
The pair  $(\Gamma, \chi)$ provides a parametrization of the phase space $\MM$, see \cite{V3}. Such reducible Riemann surfaces were introduced in
\cite{KrV}. 
\vskip 0.5in
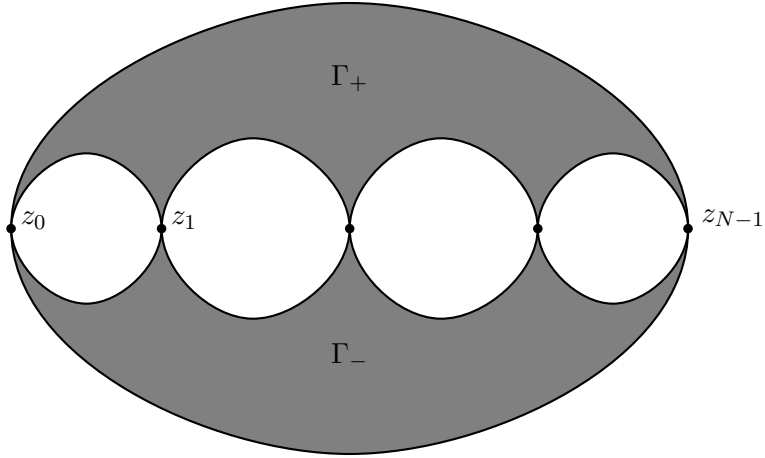
\begin{figure}[tb]
\begin{center}
\begin{pspicture}(0,0)(12,9)
\psccurve[showpoints=false,fillstyle=solid,fillcolor=gray]
(1.5,4.5) (6,7.5)  (10.5, 4.5) (6,1.5)
\psccurve[showpoints=false,fillstyle=solid,fillcolor=white]
(1.5,4.5) (2.5,5.5)  (3.5, 4.5) (2.5,3.5)
\psccurve[showpoints=false,fillstyle=solid,fillcolor=white]
(8.5,4.5) (9.5,5.5)  (10.5, 4.5) (9.5,3.5)
\psccurve[showpoints=false,fillstyle=solid,fillcolor=white]
(3.5,4.5) (4.7,5.7)  (6, 4.5) (4.7,3.3)
\psccurve[showpoints=false,fillstyle=solid,fillcolor=white]
(6,4.5) (7.2,5.7)  (8.5, 4.5) (7.2,3.3)
\psdots(1.5,4.5)(3.5,4.5)(6,4.5)(8.5, 4.5)(10.5, 4.5)
\rput(1.8,4.5){\makebox(0,0)[cb]{$z_0$}}
\rput(3.8,4.5){\makebox(0,0)[cb]{$z_1$}}
\rput(11.1,4.5){\makebox(0,0)[cb]{$z_{N-1}$}}
\rput(6,6.5){\makebox(0,0){$\Gamma_+$}}
\rput(6,2.8){\makebox(0,0){$\Gamma_-$}}


\end{pspicture}
\end{center}
\caption{\small   The reducible Riemann surface $\Gamma$ which consist of two components $\Gamma_-$ and $\Gamma_+$. These components are copies of the Riemann sphere attached to each other at the points $z_0, z_1, \hdots, z_{N-1}$. }
\label{fi:corner}
\end{figure}

\noi
Now we introduce a family of compatible Poisson brackets.
Changing  spectral variable $\l \rightarrow z=-1/\l$ we have
$$
\chi(z)=-\frac{1}{4} +\sum_{m=0}^{N-1} \frac{\rho_m'}{z_m-z},
$$
where  poles $z_m$ accumulate near the origin.
\beq\label{spec}
0< z_0< z_1<\hdots < z_{N-1}.
\eeq
The Weyl function belongs to the class $\Rat_N$ of the  rational functions on the Riemann sphere which have $N$ distinct poles  \ref{spec}  and vanish at infinity.

Consider a connected set  of  functions  $\chi(z)$  on  $\CP^1$   with the property $\chi(\infty)=0$ and $N$  simple  poles at $z_0,z_1,...,z_{N-1}$.
We denote all such functions as $\Rat_N$.
Apparently any function from $\Rat_N$ can be uniquely written as
$$
\chi(z)=-\frac{q(z)}{p(z)}, \quad\quad\text{where}\quad  p(z)= \prod_{k=0}^{N-1} (z-z_k),\quad q(z)=q_0 \prod_{k=1}^{N-1}
(z-\gamma_k).
$$
The space $\Rat_N$  has complex dimension $2N$ and  $z-q(z)$ complex coordinates
$$z_0,...,z_{N-1}; q(z_0),...,q(z_{N-1}).$$
Any such function  can be represented as
\beq\label{sf}
\chi(z)=  \sum_{k=1}^{N} \frac{\rho_k}{z_k-z} ,\quad\quad \rho_k= -\R_{z_k} \chi (z).
\eeq
We have another set of $z-\rho$ coordinates
$$z_0,...,z_{N-1}; \rho_0,...,\rho_{N-1}.$$

To introduce a formula for the hierarchy we consider a meromorphic differential $\omega_{p \, q}^{f}$ on   $\Gamma$ which depends on the  entire
function $f(z)$ and two points $p$ and $q$
\beq\label{apbt}
\omega_{p\, q}^f= \frac{ \epsilon_{p q}(z)}{p -q}\times f(z)  \chi(z) \(\chi(p)-\chi(q)\),
\eeq
where
$$
\epsilon_{p q}(z)=\frac{1}{2\pi i}\, \[\frac{ 1}{z-p } - \frac{ 1}{z-q}\] dz;
$$
is the standard differential Abelian differential of the third kind with  residues $\pm 1$ at the points $p$ and $q$.

\begin{thm}\label{JI}\cite{GV, V4}
For any entire function $f$   the Poisson bracket  \ref{apb} satisfies the Jacobi identity
$$
\{\{\chi(p), \chi(q)\},\chi(r)\}+ cp\,(p,\,q,\, r)=0.
$$
\end{thm}
Finally,  we can state the theorem
\begin{thm}\cite{V3}
The  family  Poisson brackets  for the Camassa--Holm  is produced by \ref{apb} with the differential $\omega_{p q}(z)$ given by \ref{apbt}. The infinite hierachy \ref{ihpb} is produced by \ref{apb}  for a particular choice of $f(z)=z^n, \; n=0,1,\hdots.$
\end{thm}
The map $\MM \longrightarrow (\Gamma, \chi)$ induces Poisson structure on
$\MM$.

By the Cauchy formula from \ref{apb} we have for any entire $f(z)$
\bey
\{\chi(p),\chi(q)\}^f & = & \R_p \, \omega_{p\, q}^f + \R_q \, \omega_{p\, q}^f +\R_{\infty} \, \omega_{p\, q}^f \\
                & = & \frac{f(p) \chi(p)-f(q)\chi(q)}{p-q}\(\chi(p)-\chi(q)\)+
\R_{\infty} \, \omega_{p\, q}^f.
\eey
If $f(z)=z^n,\, n=0,1,\hdots;$ then the residue at infinity vanishes identically only for $n=0$ or $1.$
When $f(z)=1$ we obtain quadratic Poisson algebra corresponding to the {\it rational} solution of CYBE or the Attiyah-Hitchin bracket
\beq\label{ahpb}
\{\chi(p),\chi(q)\}^1=\(\chi(p)-\chi(q)\)\frac{\chi(p)-\chi(q)}{p-q}.
\eeq
Another quadratic Poisson algebra is obtained for $f(z)=z$ and it corresponds to the {\it trigonometric} solution of CYBE
\beq\label{tpb}
\{\chi(p),\chi(q)\}^z=\(p\chi(p)-q\chi(q)\)\frac{ \chi(p)- \chi(q)}{p-q}.
\eeq
It can be verified directly that \ref{ahpb} and \ref{tpb}  satisfy Jacobi identity.

\section{The  open Toda lattice.}
The open finite  Toda lattice is a mechanical system of $N$--particles connected by
elastic  strings.  The Hamiltonian of the system is
$$
\HH_1=\sum\limits_{k=0}^{N-1}\frac{p_k^2}{ 2} +\sum\limits_{k=0}^{N-2}e^{q_k-q_{k+1}}.
$$
Introducing the classical Poisson bracket
\beq\label{cpb}
\{f,g\}_{\pi_0}=\sum\limits_{k=0}^{N-1} \frac{\pa f}{\pa q_k} \frac{\pa g}{ \pa p_k}-
                                \frac{\pa f}{ \pa p_k} \frac{\pa g}{\pa q_k},
\eeq
we write the equations of motion as
\begin{eqnarray}
X_1:\qquad\qquad q_k^{\bullet}&=&\{q_k,\HH_0\}= p_k, \nonumber\\
p_k^{\bullet}&=&\{p_k,\HH_0\}=-e^{q_k-q_{k+1}}+e^{q_{k-1}-q_{k}},\qquad\qquad
k=1,\ldots,N-1. \nonumber
\end{eqnarray}
We put $q_{-1}=-\infty,\;  q_N=\infty$ in all formulas. These equations define the vector field $X_1$.

Following \cite{KK},  consider  functions $\chi(\l)$ with the properties {\it i.}
analytic in the
half--planes $\Im z > 0$ and $\Im z < 0 $. {\it ii.}  $\chi(\bar{z}) = \overline{\chi(z)}$, if
$\Im z \neq 0$
{\it iii.} $\Im \chi(z) > 0, $ if $ \Im z >0$. All such function are called $R$--functions. They play
central role in the spectral theory of selfadjoint operators. The Weyl function of a Jacobi matrix is an $R$--function.

We denote by $\Rat_N'$ the subset of all functions from $\Rat_N$  which satisfy the condition
$$
q_0=\sum_{n=1}^N \rho_n=1.
$$
The Weyl functions of a Jacobi matrix are exactly those $R$--functions that belong to  $\Rat_N'$.
This implies that  all  $z_k$ are real and $\rho_k >0$  in the representation \ref{sf}.

We consider two functionals
$$
\Phi_1=I_0+I_1+\hdots + \hdots + I_{N-1},\qquad\qquad\qquad \Phi_2= \log q_0;
$$
where $I_k$    are defined by
$$
I_k=\int^{z_k}_{\infty} \frac{d \zeta} {f(\zeta)},\qquad k=0,1,\hdots, N-1.
$$
\begin{thm}\cite{V4} The family of Poisson brackets for the open Toda lattice is produced by
a Dirac restriction of  the Poisson bracket \ref{apb} with the differential \ref{apbt} on the sub-manifold
\beq\label{subman}
\Phi_1=c_1,\qquad\qquad\qquad \Phi_2=c_2;
\eeq
is given by  formula \ref{apb} where the  new modified differential $\tilde{\omega}_{p\, q}^f$  is
$$
\tilde{\omega}_{p\, q}^f= \frac{ \epsilon_{p q}(z)}{p -q}\times f(z)  \chi(z) \(\chi(p)-\chi(q)\)-
 \epsilon_{p q}(z)\times f(z)  \chi(z) \chi(p)\chi(q)  e^{-c_2}.
$$
The infinite hierarchy of Poisson brackets \ref{ihpb} corresponds to the choice of $f(z)=z^n,\; n=0,1,\hdots.$
\end{thm}

\section{The periodic solutions of the KdV equation.}

We consider the hierarchy of flows Korteveg de Vries equation\footnote{Prime $'$  signifies the derivative in the   variable $x$ and dot $\bullet$ the derivative with respect to time.  }
$$
u^{\bullet}= \frac{3}{2}uu' -\frac{1}{4} u''',\qquad\qquad\qquad\qquad u=u(x,t);
$$
in the space $\MM$ of all infinitely differentiable $2l$-periodic functions $u(x,t)=u(x+2l,t)$.
The KdV equation is a compatibility condition for the Lax representation
$$
L^{\bullet}=\[L, A \].
$$
Where $L$ is  the Shr\"{o}dinger operator
\beq\label{sho}
L=-\partial_{x}^2+ u.
\eeq
and
$$
A=4\partial_{x}^3  - 6u\partial_x - 3u_x.
$$
The KdV equation is one in the infinite  hierarchy of vector fields
\bey
&X_0&:\qquad\qquad u^\bullet=u',\\
&X_1&:\qquad\qquad u^\bullet=\frac{3}{2}uu' -\frac{1}{4} u''',\\
&X_2&:\qquad\qquad u^\bullet=\frac{1}{16} u^{(V)}-\frac{5}{4} u'u''-\frac{5}{8}u u'''+\frac{15}{8}u^2u', \qquad\qquad etc.
\eey
Each vector field $X_k,\; k=0, 1,2,...;$ is a  compatibility condition for the commutator formalism $L^{\bullet}=\[L, A_k \]$ with some $A_k$ and $A_1=A$.
These vector fields produce an infinite hierarchy of commutative  flows $e^{tX_m},\; m=0,1,2,\ldots$.

 Gardner  and Zakharov and Faddeev   found  that each flow $X_n$
can be written as a Hamiltonian system
$$
u^{\bullet}=\{u,\HH_n\}_{\pi_0},
$$
with  the bracket
\beq\label{gfz}
\{A,B\}_{\pi_0}=\int \frac{\d A}{\d u(x)} D \frac{\d B}{\d u(x)}\, dx
\eeq
where $D=\partial_x$ and  corresponding Hamiltonian $\HH_n$. Here the first three Hamiltonians
\bey
\HH_0&=&\frac{1}{2}\int  u^2 \,dx,\\
\HH_1&=&\frac{1}{4}\int \[ u^3  + \frac{1}{2} u'^{\,2}\] \,dx,\\
\HH_2&=& \frac{1}{16}\int \[ \frac{1}{2}  u''^{\,2} +5 u   u'^{\,2}+ \frac{5}{2}u^4\]  \,dx,\qquad etc.
\eey
For each  Hamiltonian
$$
\HH_n=\int L_n(u(x),u'(x),u''(x),\hdots, u^{(n)}(x))\, dx
$$ the variational derivative can be computed by the formula
$$
\frac{\d \HH_n}{\d u(x)} =\sum_{k=0}^{n} (-1)^k\;  \partial_x^{\,k} \; \frac{\partial L_n}{\partial u^{(k)}(x)}.
$$
The Hamiltonians commute with respect to the Poisson bracket. Thus we have a hierarchy of commutative Hamiltonian flows.

For the periodic potentials the spectral curve  is defined by
$$
\Gamma=\{Q=(z,w)\in { \C}^{2}:\quad   R(z, w)=\det\[ w I-T(z)\]= w^{2}- 2 w \Delta(z)  + 1=0\},
$$
where $2\Delta(z)=\tr T (z),$    and represents  two  sheets  covering of $\CP^1$.
In other words on $\Gamma$ there exists two functions $z=z(Q)$ and $w=w(Q)$ which determine an embedding of the spectral curve into $\C^2$.
In this case the two differentials are $dE=dz$ and $dp=\frac{1}{i} d\, log w$.

The key role in our considerations is played by  the scalar Weyl function $\chi(x,Q)$ defined the formula
\beq\label{kdvweyl}
i\chi(x,Q)=\frac{\p}{\p x} \log e(x,y,Q).
\eeq
where $e(x,y,Q)$ is the standard Floquet solution normalized as $e(y,y,Q)=1$
and
\beq\label{chic}
i\chi(x,Q)=\frac{w(Q)-T_{11}(y,z)}{ T_{12}(y,z)}=\frac{ T_{21}(y,z)}{w(Q)- T_{22}(y,z)}.
\eeq
Now we defined  all objects that constitute a quadruples $\{\Gamma, dp, dE, \chi\}$.

First we define the   meromorphic differential  $\epsilon_{QP}(R)$ is  of the third kind on $\Gamma$  with poles  at the points $Q=(z_Q,w_Q)$ and $P=(z_P,w_P)$  and the residues there equal to  $\pm 1$
$$
\epsilon_{QP}(R)=\(\frac{w(R)-w^{-1}(Q)}{z(R) -z(Q)}- \frac{w(R)-w^{-1}(P)}{z(R) -z(P)}\)\frac{d z(R) }{ w(R)-w^{-1}(R)}.
$$
This differential has  simple  poles at the points $P$ and $Q$. The residues are $\pm 1$. Also singularities may arise from the factor
\beq\label{dfspec}
\frac{d z(R) }{ w(R)-w^{-1}(R)}
\eeq
 when $w(R)=\pm 1$. This happens at the points  $z_k^-  \leq   z_k^+$ of the periodic/anti periodic spectrum.  If the spectrum is simple $z_k^-  <   z_k^+$ then at these branch points  the differential $dz(R)$   has  simple zeros  and these zeros annihilate the singularity of the denominator. If $z_k^-  =    z_k^+$,     then the singularity is  nodal. The  function $z$ serves as a local parameter in the vicinity of these points and the differential $dz(R)$ does not vanish.  The differential \ref{dfspec} changes sign under involution $\tau_p$ permuting
sheets of the curve and it has residues of opposite sign on different sheets.

The  meromorphic differential  $\omega_{QP}^f$ is such that
\beq\label{fff}
\omega_{QP}^f(R)= \frac{1} {2\pi i}
f(z(R))\times \epsilon_{QP}(R)\times  \chi(R)(\chi(Q)-\chi(P))\times \frac{\Omega(Q)+ \Omega(P)}{2},
\eeq
where the function  $f(z)$ is an entire function, {\it e.g.} $f(z)=z^n,\; n=0,1,2,...$.

The deformation factor $\Omega(Q)$ is  defined by
$$
\Omega(Q)=\frac{w(Q)+ w^{-1}(Q)}{w(Q)-w^{-1}(Q)}.
$$
It becomes infinite at the branch and intersection points of the curve $\G$.

It is easy to state analytic properties of \ref{fff}. Fix two  points $Q$ and $P$  away from the branch or intersection points of $\G$.
The differential \ref{fff} has poles at the points $M_k=M_k(x)$  coming from $\chi(R)=\chi(x,R)$    and
it also has poles  at the points $P$ and $Q$ and also at the points of the double spectrum (unopened zones)  $(z_k^\pm,  (-1)^k)$.  The last group of poles arises
from $\epsilon_{QP}(R)$.  The residues of
\ref{fff} on different sheets of the curve above the double spectrum are of the opposite sign and cancel each other if we integrate over the contour containing  these points.

{\bf Conjecture.} \cite{V5}. {\it The family of Poisson brackets for KdV hierarchy is produced by  formula \ref{apb} with the-one form $\omega_{QP}^f$ defined by
\ref{fff}. The hierarchy of Poisson brackets \ref{ihpb}  is produced for a particular choice of $f(Q)=z(Q)^n,\; n=0,1\hdots.$}

Now we explain why we think our conjecture is true. The formula is modeled upon the simplest formula \ref{apbt} for the CH hierarchy. Moreover, for a particular choice of $f(z)=1$ by the residue theorem the conjectured formula   produces the quadratic algebra for the Gardner-Zakharov-Faddeev bracket
\beq\label{rpb}
\{\chi(P), \chi(Q)\}=\frac{\( \chi(P)  -  \chi(Q)  \)^2}{z(P)-z(Q)}   \times \frac{\Omega(P)+ \Omega(Q)}{2}.
\eeq
The last formula was proved  by direct computations for the first time in \cite{V2}. It corresponds to the rational solution for the PB on the entries of the monodromy matrix and we call it in \cite{V2} the deformed Attiah--Hitchin bracket, compare \ref{ahpb}.

\section {The Poisson brackets associated with differentials of the second kind.}

In all previous examples the one-form $\omega_{PQ}$  was constructed using differential of the third kind. Now we want to show an example
of the Poisson bracket and the  one form associated with a differential of the second kind, \cite{V6}.

We consider the space $Rat_N$ of the meromorphic functions on the Riemann  sphere.
Let
$$
\epsilon_P^{(n+1)}(z)= \frac{n!}{2\pi i}\frac{ dz}{(z-z_P)^{(n+1)} },  \qquad\qquad\qquad \qquad n=1,2,\hdots;
$$
be Abelian differential of the second kind with a pole of degree $n+1$ at the point $p$. Note that for any function $f(z)$ which is
holomorphic in the vicinity  of the point $P$ we have
$$
\int\limits_{\overset{\curvearrowleft}{O_P}} \epsilon_P^{(n+1)}(z)  f(z) = f^{(n)}(P).
$$
Let us define  the new meromorphic 1-form
$$
\omega_{PQ}^f= \epsilon^{(2)}_P \times \chi(z) f(z)\chi(Q) - \epsilon^{(2)}_Q \times \chi(z) f(z)\chi(z_P).
$$
The new analytic Poisson brackets are defined by usual formula \ref{apb}.
\noi

When $f(z)=1$ or $z$   we can obtain closed expression in terms of $\chi$ and its first derivative.
For $f(z)=1$ we have
$$
\{\chi(P),\chi(qQ)\}^1= \chi'(P) \chi(Q) - \chi'(Q) \chi(P).
$$
For $f(z)=z$ we  have
$$
\{\chi(P),\chi(Q)\}^z= z_P\chi'(P) \chi(Q) - z_Q\chi'(Q) \chi(P).
$$
It can be proved that these formulas define genuine Poisson brackets {\it e.g.} they satisfy the Jacobi identity.


\newpage
\begin{thebibliography}{***}

\bibitem{AH}  Attiyh  M. and Hitchin N. The geometry and dynamics of magnetic monopoles. M. B. Porter Lectures. Princeton University Press, Princeton,
NJ, 1988. viii+134 pp.



\bibitem{FT} Faddeev and Takhtadzian  L. Hamiltonian methods in the Theory of Solitons. Springer. 1986

\bibitem{G} Gardner, Clifford S. {\it  Korteweg-de Vries equation and generalizations. IV. The Korteweg-de Vries equation as a Hamiltonian system.}
 J. Mathematical Phys. 12 1971 1548–-1551.


\bibitem{GK} F.R.  Gantmacher and M.G.  Krein
{\it Oscillation Matrices and Small Oscillations of Mechanical Systems.} (Russian)
Gostekhizdat, Moscow--Leningrad. (1941).




\bibitem{FGEL} Gelfand I.  and   Fokas A.,  Quadratic Poisson algebras and their infinite-dimensional extensions. J. Math. Phys. 35,  1994, no. 6, 3117--3131.

\bibitem{DM} H. Dym and H. McKean
{\it Gaussian Processes, Function Theory, and the Inverse  Spectral Problem.}
Academic Press, New York, Sun Francisco, London. (1976).

\bibitem{KK2}  I.S. Kac and M.G. Krein
{\it  On the spectral function of the string.}
 Transl. Amer. Math. Soc. v.  103, pp. 19--102, (1974).

\bibitem{GD} Gelfand I.M., and Dikii, The resolvent and Hamiltonian systems, Funct. Anal.
Appl. 11 (2) (1977), pp. 93--105.

\bibitem{FG} Gekhtman M, and Faybusovich L., Poisson brackets on rational functions and multi-Hamiltonian structure for integrable lattices. Phys. Lett. A 272 2000, no. 4, 236--244.

\bibitem{BD}  Belavin A.,  and Drinfeld V., Solutions of the classical Yang-Baxter equation for simple Lie algebras. (Russian) Funktsional. Anal. i Prilozhen. 16 1982,
no. 3, 1--29, 96.

\bibitem {KK}  Kac I. and   Krein M. 1974.
{\it R functions -- analytic functions mapping the upper half--plane into itself},
Amer Math Soc. Transl, 103, pp.  1--19.


\bibitem{KP}  Krichever, I. M.; Phong, D. H. Symplectic forms in the theory of solitons. Surveys in differential geometry: integrable systems, 239--313,
Surv. Differ. Geom., IV, Int. Press, Boston, MA, 1998.

\bibitem{KrV} Krichever, I.; Vaninsky, K. L. The periodic and open Toda lattice. Mirror symmetry, IV (Montreal, QC, 2000), 139--158, AMS/IP Stud. Adv. Math., 33, Amer. Math. Soc., Providence, RI, 2002.

\bibitem{MA} Magri F. A simple model of the integrable Hamiltonian equation. Journal of Mathematical Physics, Volume 19, Issue 5, pp. 1156--1162 (1978).
\bibitem{GV} Gekhtman, M. I.; Vaninsky, K. L. The family of analytic Poisson brackets for the Camassa-Holm hierarchy. Math. Res. Lett. 15 (2008), no. 5, pp. 867--879.

\bibitem{MaV}  Malamud M.M. and K.L. Vaninsky.
{\it paper in preparation}.



\bibitem{MCV2} McKean H. and Vaninsky K. Action-Angle Variables
for the Cubic Schrödinger Equation.  Communications on Pure and Applied Mathematics, Vol. L, 0489–-0562 (1997)


\bibitem{MT} McKean H. and Trubowitz. E. Hill’s operator and hyperelliptic function theory in the
presence of infinitely many branch points, Comm. Pure Appl. Math. 29, 1976, pp. 143–-226.


\bibitem{V1} Vaninsky, K. L. The Atiyah-Hitchin bracket and the open Toda lattice. J. Geom. Phys. 46 (2003), no. 3-4, 283--307.
\bibitem{V2}  Vaninsky, K. L. The Atiyah-Hitchin bracket and the cubic nonlinear Schrodinger equation. IMRP Int. Math. Res. Pap. 2006, 17683, pp. 1--60.
\bibitem{V3} Vaninsky,  K. L. Equations of Camassa-Holm type and Jacobi ellipsoidal coordinates. Comm. Pure Appl. Math. 58 (2005), no. 9, 1149–1187.
\bibitem{V4} Vaninsky,  K. L.    The  hierarchy  of Poisson brackets   for the open Toda lattice  and its' spectral curves. {\it To appear}.
\bibitem{V5} Vaninsky, K.L. The hiearchy of Poisson brackets for the periodic KdV and its spectral curve. {\it paper in preparation.}
\bibitem{V6} Vaninsky, K.L. The hiearchy of strange Poisson brackets and differentials of the second type. {\it paper in preparation.}


\bibitem{PT}   Pastur, L. A.; Tkachenko, V. A. Spectral theory of a class of one-dimensional Schrödinger operators with limit-periodic potentials. (Russian) Trudy Moskov. Mat. Obshch. 51 (1988), 114--168, 258; translation in Trans. Moscow Math. Soc. 1988, 115–166




\bibitem{Kr1} Krichever I.  Spectral theory of two-dimensional periodic operators and its applications. (Russian) Uspekhi Mat. Nauk 44; 1989, no. 2(266), 121--184;







\bibitem{KR} Krichever, I.M. Spectral theory of two-dimensional periodic operators and its applications. (Russian) Uspekhi Mat. Nauk 44 (1989), no. 2(266), 121--184;

\bibitem{KV} Krichever  I.  and Vaninsky K., The periodic and open Toda lattice. (English summary) Mirror symmetry, IV (Montreal, QC, 2000), 139–-158,
AMS/IP Stud. Adv. Math., 33, Amer. Math. Soc., Providence, RI, 2002.





\bibitem{W}   H. Weyl
{\it   Uber gewohnliche Differentialgleichungen mit Singulariten und die
zugehorigen Entwicklungen willkurlichen Funktionen.}
Math Ann,  68, 1910,  pp. 220--269.

\bibitem{ZF} Zakharov V.E.  and Faddeev L.D. {\it Korteweg—de Vries equation: A completely integrable Hamiltonian system.} Funct. Anal. Appl. 1971. Vol. 5. P. 280-—287.

\bibitem{NV}  Novikov S.  and Veselov A. Poisson brackets and complex tori. Algebraic geometry and applications. Trudy Mat. Inst. Steklov 165 (1984), 49--61.















\end{thebibliography}
\end{document}